\documentclass[aps,prl,showpacs,reprint]{revtex4-1}
\usepackage{dcolumn,graphicx,color}
\usepackage{amsfonts,amsmath,amssymb}
\usepackage{bm}

\newcommand{\br}{\mathbf{r}}
\newcommand{\bd}{\mathbf{d}}
\newcommand{\bOme}{\bm{\Omega}}
\newcommand{\bome}{\bm{\omega}}
\newcommand{\bF}{\mathbf{F}}
\newcommand{\bT}{\mathbf{T}}
\newcommand{\bM}{\mathbf{M}}
\newcommand{\bq}{\mathbf{q}}
\newcommand{\rmd}{\mathrm{d}}
\newcommand{\Lrft}{L_\mathrm{RFT}}
\newcommand{\Lsbt}{L_\mathrm{SBT}}
\newcommand{\Lim}{L_\mathrm{image}}
\newcommand{\Sp}{\mathrm{Sp}}

\begin{document}
\title{Kinematics of the most efficient cilium}
\author{Christophe Eloy}
\email{eloy@irphe.univ-mrs.fr}
\altaffiliation[Permanent address: ]{Aix--Marseille University, IRPHE, CNRS, Marseille, France.}
\author{Eric Lauga}
\email{elauga@ucsd.edu}
\affiliation{Department of Mechanical and Aerospace Engineering, University of California San Diego, \\ 9500 Gilman Drive, La Jolla CA 92093-0411, USA}
\date{\today}                                           
\begin{abstract} 
In a variety of biological processes, eukaryotic cells use cilia to transport flow. Although cilia have a remarkably conserved internal molecular structure, experimental observations report very diverse kinematics. To address this diversity, we determine numerically the kinematics and energetics of the most efficient cilium. Specifically, we compute the time-periodic deformation of a wall-bound elastic filament  leading to transport of a surrounding fluid at minimum energetic cost, where the cost is taken to be the positive work done by all internal molecular motors. The optimal kinematics are found to strongly  depend on the cilium bending rigidity through a single dimensionless number, the Sperm number, and closely resemble the two-stroke ciliary beating pattern observed experimentally.
\end{abstract}
\maketitle

Cilia are slender filaments, typically a few microns in length, used by eukaryotic cells to transport or sense flows \cite{Sleigh1974,Lauga2009}.  Familiar examples include those  densely packed on the surface of \emph{Paramecia} enabling  locomotion \cite{Sleigh1974}, cilia covering our  airways and helping expel mucus towards the pharynx \cite{Sleigh1988}, or those responsible of the left-right symmetry  breaking during embryonic development \cite{Hirokawa2009}. 

The cilium internal structure has been highly conserved throughout evolution. It generally consists of a central pair of microtubules surrounded by nine microtubule doublets, forming the so-called `$9+2$' structure \cite{Satir2007}. The deformation of the cilium is achieved by the action of
ATP-fueled protein motors (dynein) able to generate internal moments from the relative sliding of adjacent microtubule doublets.  Yet, the mechanisms that regulate dynein activity and thus  ciliary deformation  are not well understood \cite{Lindemann2010}. 

\begin{figure}[b] 
\includegraphics[scale=0.24]{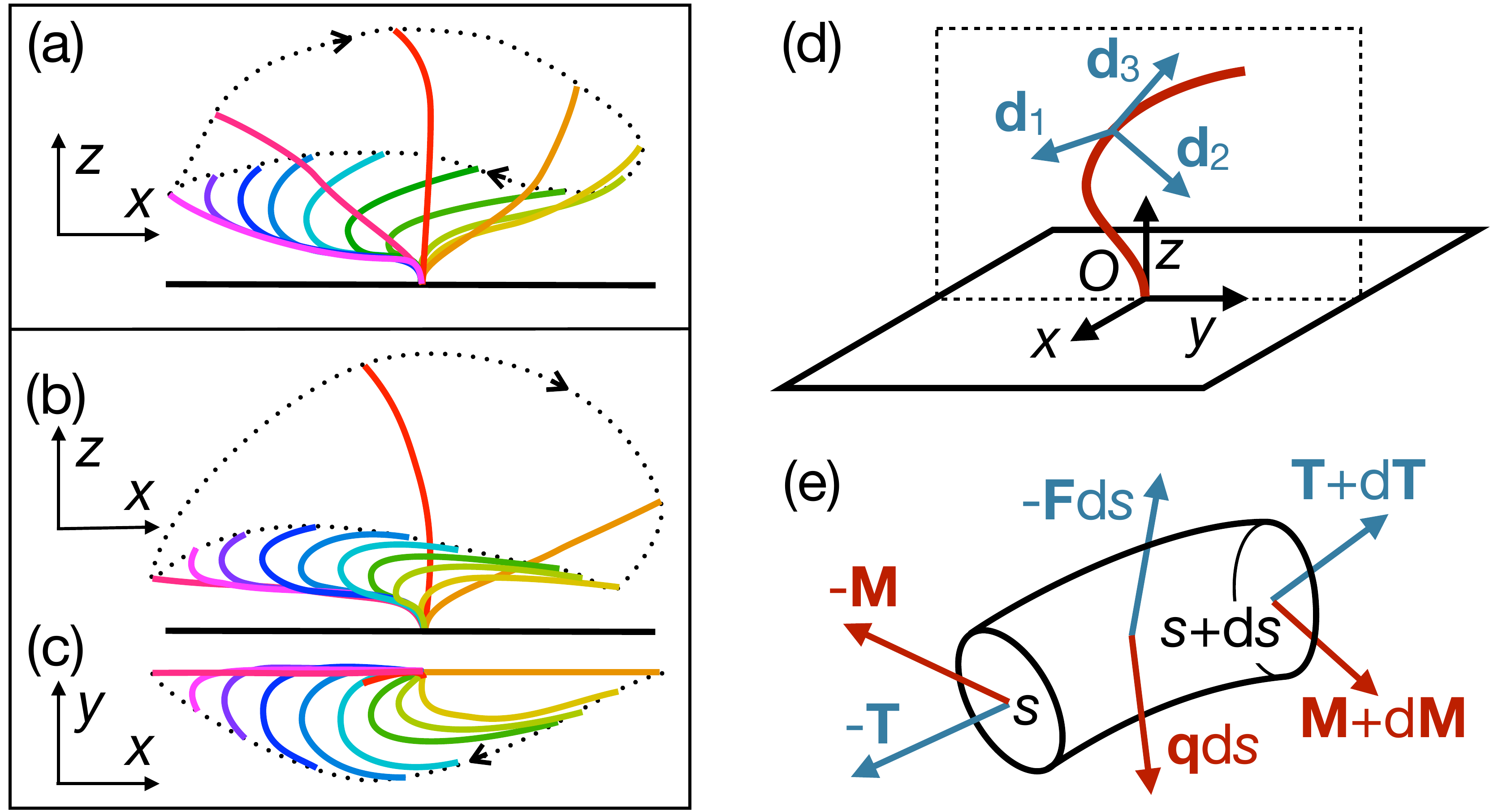} 
\caption{(color online) Typical two-dimensional (a) and three-dimensional (b,c)  cilium kinematics as observed for \emph{Pleurobrachia} \cite{Sleigh1968} and \emph{Mytilus edulis} \cite{Aiello1972} respectively. Sketch of the coordinates (d) and of the forces and moments that apply on a filament element $\rmd s$ (e).}
\label{fig:sketch}
\end{figure}

The beating cycle of a cilium typically consists of two phases (Fig.~\ref{fig:sketch}, left): an effective stroke aimed at generating flow and during which the cilium is almost straight while moving in a plane normal to the cell surface, followed by a recovery stroke during which the cilium returns to its initial position by exhibiting large curvatures and possibly  moving out of the normal plane.  Past experimental works have shown that cilia from different cells can exhibit qualitatively different kinematics \cite{Sleigh1974}. However, the parameters,  physical or biological, that select or constrain these kinematics are still unknown.

In this Letter, we address this open question by computing the optimal kinematics of an elastic cilium attached to a wall, 
i.e.~the time-varying deformation that minimizes the energetic cost for a given transport of the surrounding fluid.  Recent work focused on the optimal deformation of flagellated cells  \cite{Tam2011} and cilia \cite{Osterman2011} by minimizing the  energy lost to viscous dissipation in the fluid. Here we argue that one needs to rigorously consider the internal structure of the cilia and measure the energetic costs as the sum of the positive works done by the internal  molecular moments, similarly to current models of muscle energetics \cite{Alexander92}.  This modeling approach leads to  optimal ciliary kinematics  displaying the experimentally-observed two-stroke cycle and  strongly dependent on the cilium bending rigidity.

The cilium is modeled as an inextensible elastic filament of length $L$ and radius $a$, clamped normally into the plane $Oxy$ (Fig.~\ref{fig:sketch}). 
The filament centerline is described by the vector $\br(s,t)$, where $s$ is the curvilinear coordinate, and the material frame $(\bd_1,\bd_2,\bd_3)$ describes the local orientation of the filament such that
\begin{equation}
\bd_i' = \bOme \times \bd_i, \quad
\dot\bd_i = \bome \times \bd_i, \quad
\mbox{for $i=1\cdots3$},
\end{equation}
where primes and dots note differentiation with respect to $s$ and $t$ respectively, $\bOme$ is the Darboux vector \cite{Powers2010,Audoly2010}, and $\bome$ the angular velocity.

The balance of forces and moments on a cross-section are expressed by the Kirchhoff equations for a rod \cite{Audoly2010}
\begin{equation}\label{eq:Kirchhoff}
\bT' - \bF = 0, \quad
\bM' + \bd_3\times\bT + \bq =0, 
\end{equation}
 where $\bT$ and $\bM$ are the internal tension and elastic moment respectively, $\bq$ is the internally applied moment per unit length, and $-\bF$ is the drag per unit length (the opposite of the force $\bF$ exerted by the filament on the surrounding fluid). 
The internal moment $\bq$ models the discrete distribution of forces generated by the dynein arms. Since  these forces can only induce sliding of adjacent microtubule doublets, the energy needed to produce torsion is of order $L/a$ larger than that necessary to produce bending \cite{Hines1985}. In the slender limit, $a/L\ll 1$, relevant to biology we can thus assume that there is no twist and no tangential component of the internal moment, $\bOme\cdot\bd_3=\bq\cdot\bd_3=0$.

The Hookean constitutive relation relates the elastic moment $\bM$ to the Darboux vector $\bOme$ through a linear relation. In the absence of torsion and for an axisymmetric filament it simplifies to
$\bM = B\, \bOme = B\, \bd_3\times\bd_3'$, 
with $B$ the bending rigidity. Combining this constitutive relation and Eq.~(\ref{eq:Kirchhoff}), the internal moment $\bq$ can be expressed as a function of $\bd_3$ and $\bF$ alone
\begin{equation}\label{eq:q}
\bq = B \, \bd_3''\times\bd_3 + \int_s^L \bF(\xi)\rmd\xi.
\end{equation}

Assuming  the cilium kinematics  known, we need to evaluate the hydrodynamic forces $\bF$ in order to fully determine the internal moment  $\bq$. Since cilia are few microns long, their Reynolds number is small and the surrounding flow follows the Stokes equations. In this limit, $\bF(s)$ represents a distribution of stokeslets, which are the Green functions (point forces) of the Stokes equations. 

Taking advantage of the small filament aspect ratio, we use slender-body theory \cite{Johnson1980,Gotz2000}, which allows to relate $\bF(s)$ to the instantaneous distribution of velocities, $\dot\br(s) $, along the cilium  centerline. To take into account the presence of the no-slip wall to which the cilium is anchored, slender-body theory is supplemented with Blake's system of hydrodynamic images \cite{Blake1971} allowing to  write
\begin{equation}\label{eq:SBT+images}
\dot\br (s)= \Lrft\cdot \bF + \Lsbt(\bF) +\Lim (\bF),
\end{equation}
where $\Lrft$ is the local linear operator of so-called {resistive-force theory} \cite{Lauga2009} given by $\Lrft  = ({\bf 1}+\bd_3 \bd_3)/\xi_\perp$,
with ${\bf 1}$ the $3\times 3$ identity  matrix, $\xi_\perp=4\pi\mu/\ln(L/a)$, and $\mu$ the dynamic viscosity of the fluid.  The linear integral operators $\Lsbt$ and $\Lim$, which are of order $\ln(L/a)$ smaller,  account for the cilium--cilium and cilium--wall hydrodynamic interactions. Their full expressions can be found  in Refs.~\cite{Johnson1980} and \cite{Blake1971}.  Numerically, Eq.~(\ref{eq:SBT+images}) is  regularized by using Legendre polynomials to diagonalize the singular part of $\Lsbt$ \cite{Gotz2000}. Once this regularization is performed, the discretization and inversion of Eq.~(\ref{eq:SBT+images}) is straightforward. The resulting computational implementation of Eq.~(\ref{eq:SBT+images}) is correct to order 
 $O(a/L)$.

Without loss of generality, the net flow transported by the cilium is assumed to occur in the $x$-direction. This transport is quantified by the flow rate, $Q$,  across the  $Oyz$ half-plane (Fig.~\ref{fig:sketch}), which  is equal, by virtue of incompressibility, to the flow rate trough any parallel half-plane, and can thus  be  evaluated  in the far field for convenience \cite{Osterman2011}.  Far from the filament, the flow is dominated by the contribution of the stokeslets along the cilium, $\bF(s)$, and their  images, which consist of stokeslets, force dipoles, and source dipoles \cite{Blake1971}. Combined together, these singularities are equivalent at leading order to a symmetric combination of two force dipoles located in $O$,  known as a stresslet \cite{Batchelor1970_2}, leading to a flow rate given by
\begin{equation}\label{eq:Q}
Q = \frac{1}{\pi \mu} \langle \int_0^L F_x r_z \,\rmd s \rangle,
\end{equation}
where brackets denote time-averaging. 

The power, $P$, expended internally to transport  flow is equal to the power consumed by the internal moments, $\bq$,  where only the positive works are accounted for \cite{Alexander92}
\begin{equation}
P = \langle \int_0^L \max\left(0,\bq\cdot \bome\right) \, \rmd s \rangle.
\end{equation}
Distinguishing between positive and negative work means  that the dynein arms cannot harvest energy,  which breaks the conservative nature of elastic energy. It results that the mean power spent by the internal moment is  larger than the power given to the fluid, i.e.
$
P \ge \langle \int_0^L \bF\cdot \dot\br \, \rmd s \rangle.
$

From the definition of the flow rate, $Q$, and the mean power, $P$, a dimensionless efficiency can be constructed similarly to the one proposed in Ref.~\cite{Osterman2011} as
\begin{equation}\label{eq:eta}
\xi = {Q^2 \mu^2}/({P\, \xi_\perp L^3}).
\end{equation}
With this definition, the transport efficiency, $\xi$, does not depend on the beat angular frequency, $\omega$, nor on the aspect ratio at first order, since the flow rate scales as $Q\sim \omega L^3\xi_\perp/\mu$ and the power as $P \sim \xi_\perp \omega^2 L^3$.

 \begin{figure}[b] 
\includegraphics[width=0.48\textwidth]{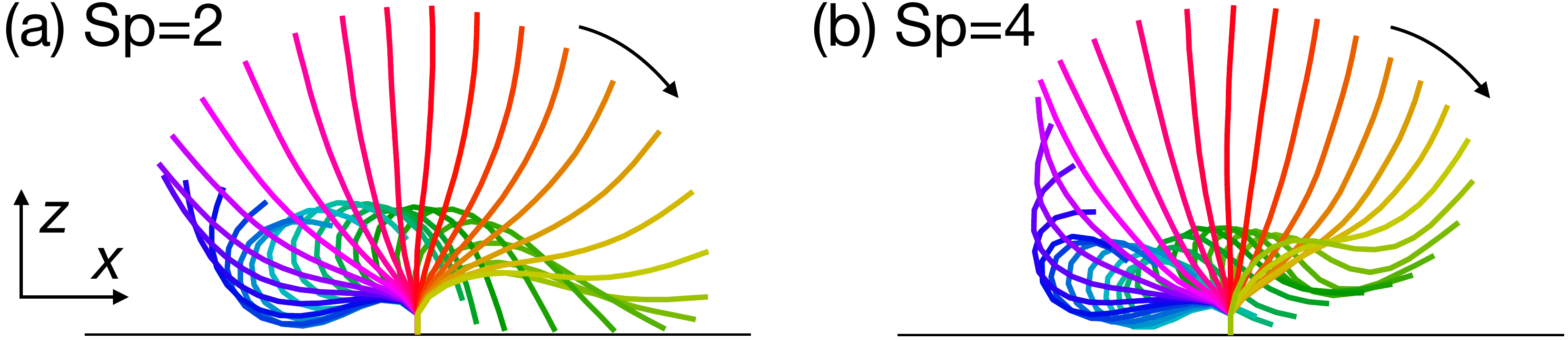} 
\caption{(color online) Two-dimensional optimal ciliary motions for $\Sp =2$ (a) and $\Sp=4$ (b). The stroboscopic views show the cilium every 1/32th of period.}
\label{fig:2D}
\end{figure}

\begin{figure*}[t] 
\includegraphics[width=0.97\textwidth]{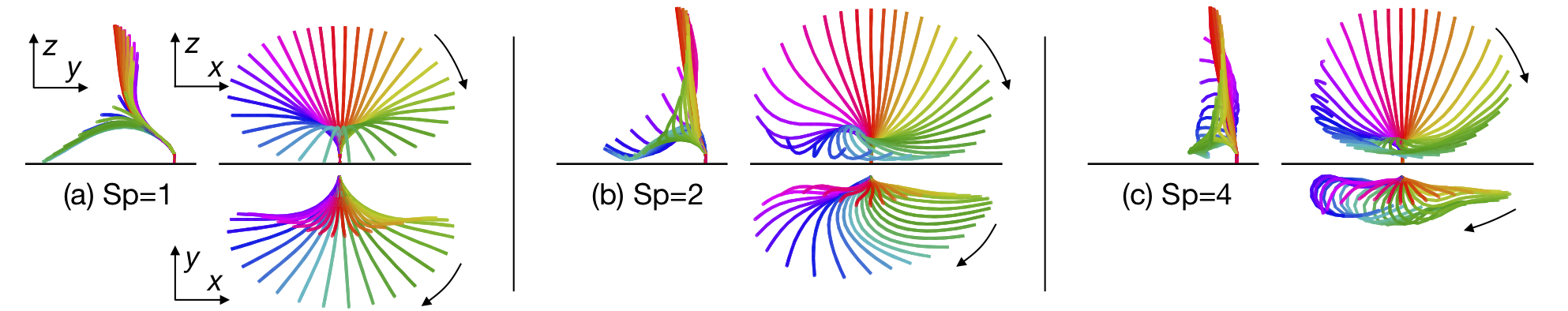} 
\caption{(color online) Three-dimensional optimal ciliary motions for $\Sp =1$ (a), $\Sp =2$ (b),  and $\Sp=4$ (c).}
\label{fig:3D}
\end{figure*}

Dimensional analysis shows that the problem is entirely governed by two dimensionless numbers. The first is  the aspect ratio of the cilium, $L/a$. The cilia covering the body of \emph{Paramecium} have $L\approx12\,\mu$m and $a\approx0.12\,\mu$m, so we will assume $L/a=100$. Note that the aspect ratio appearing only logarithmically in the problem through Eq.~(\ref{eq:SBT+images}), its influence is  essentially negligible. The second dimensionless number is the Sperm number, defined as
\begin{equation}\label{eq:Sperm}
\Sp = L \left({\omega \xi_\perp}/{B}\right)^{{1}/{4}},
\end{equation}
which measures the ratio between the cilium length and the elasto-viscous penetration length. The kinematics of optimal ciliary motion depends strongly on the value of $\Sp$. For \emph{Paramecium},  the angular frequency is $\omega\approx 200\,$rad$\,$s$^{-1}$ \cite{Sleigh1974}, and with a bending rigidity estimated to be $B=25\,$pN$\,\mu$m$^2$ \cite{Hines1983}, one obtains  $\Sp\approx 4.6$ in water. Some cilia are shorter, such as the nodal cilia involved in embryonic development with $L \approx 2.5\,\mu$m for mice \cite{Cartwright2004} and $L \approx 5\,\mu$m for humans \cite{Hirokawa2009},  corresponding  to $\Sp\approx 1$ and $\Sp\approx 2$ respectively. Other cilia can be much longer, reaching hundreds of microns and thus $\Sp>100$, such as the cilium of \emph{Pleurobrachia} reproduced in Fig.~\ref{fig:sketch}(a).  In this Letter, we will focus on the range $1\le\Sp\le 7$, corresponding approximately to cilia of lengths $2.5 \lesssim L\lesssim 18\,\mu$m, for which the most drastic changes appear in the optimal kinematics.

For given values of both $L/a$ and $\Sp$, the  cilium kinematics maximizing the pumping efficiency, $\xi$, is computed numerically. The elastic filament is first discretized   as an assembly of $n_s$ discrete rods connected by springs \cite{Bergou2008}, and  
 the stokeslet distribution is evaluated $n_t$ times per cycle \footnote{The values $n_s=16$ and $n_t=32$ have been used in this study and allow to evaluate the efficiency of a given kinematics rapidly without compromising precision.}.  The filament  kinematics is parametrized by imposing the curvatures at $N_s$ points along the filament centerline, $N_t$ times per period. The curvatures on the $n_s\times n_t$ points are then interpolated with a cubic spline from those $N_s\times N_t$ points. Our results   are obtained with $N_s=N_t=6$, giving 36 and 72 degrees of freedom in  two and three dimensions, and the optimal kinematics are computed using a sequential programming (SQP) algorithm.  Our numerical approach is validated by comparing with the results of Ref.~\cite{Osterman2011} obtained with a bead model and an energetic measure of the dissipation in the fluid only. The transport efficiency in Ref.~\cite{Osterman2011} is $\xi=0.35\%$, corresponding here to large values of $\Sp$ and  comparing well with our optimum for  $\Sp=7$, $\xi=0.33\%$.

The optimal two-dimensional kinematics are displayed in Fig.~\ref{fig:2D} for $\Sp=2$ and 4. We see that our optimization approach, which  rigorously quantifies the internal work expended by the molecular motors,  leads to kinematics with the experimentally-observed two-stroke cycle: an effective stroke during which the filament is rotating almost rigidly around its anchor point, and a recovery stroke exhibiting large curvatures. For small values of $\Sp$ (small, or stiff, cilia), the curvature is essentially always of the same sign  (Fig.~\ref{fig:2D}a), whereas for larger values of $\Sp$ (long, or flexible, cilia), curvatures are larger and occasionally  change sign (Fig.~\ref{fig:2D}b). In order to minimize the back flow, Eq.~(\ref{eq:Q}), during the recovery stroke, the trajectory has to be as close to the wall as possible, and in order to achieve such a trajectory, large curvatures with high energetic costs are necessary; the resulting optimum is thus a balance, tuned by the value of $\Sp$, between distance to the wall and curvature.

The optimal three-dimensional ciliary kinematics have also been determined. The results are illustrated in Fig.~\ref{fig:3D} for $\Sp=1$, 2 and 4.  For small values of $\Sp$, the cilium is  rotating around an axis inclined by an angle of  approximately 45 degrees with respect to the  surface normal (Fig.~\ref{fig:3D}a). This optimal motion is similar to the observed trajectories of nodal cilia  \cite{Hirokawa2009}. For larger  values of $\Sp$, the optimal cilium kinematics breaks the $x\to -x$ symmetry \footnote{The value of the efficiency is unchanged by the transformations $x\to -x$, $y\to -y$, or $t\to -t$, and thus the recovery stroke could take place in the opposite direction. Cilium chirality constrain however the motion and it is usually observed to be clockwise.}, and, as in two-dimensions, the motion can be  decomposed into an effective stroke in the vertical plane and a recovery stroke with large curvatures. During the recovery stroke, the filament takes advantage of the third dimension to achieve a trajectory closer to the wall, and therefore more efficient than in the two-dimensional case. These three-dimensional optimal kinematics reproduce the experimental observations of real cilia motions, as can be seen by comparing the kinematics of \emph{Mytilus edulis} (Fig.~\ref{fig:sketch}b and c) with Fig.~\ref{fig:3D}c, for instance.

\begin{figure}[t] 
\includegraphics[width=0.4\textwidth]{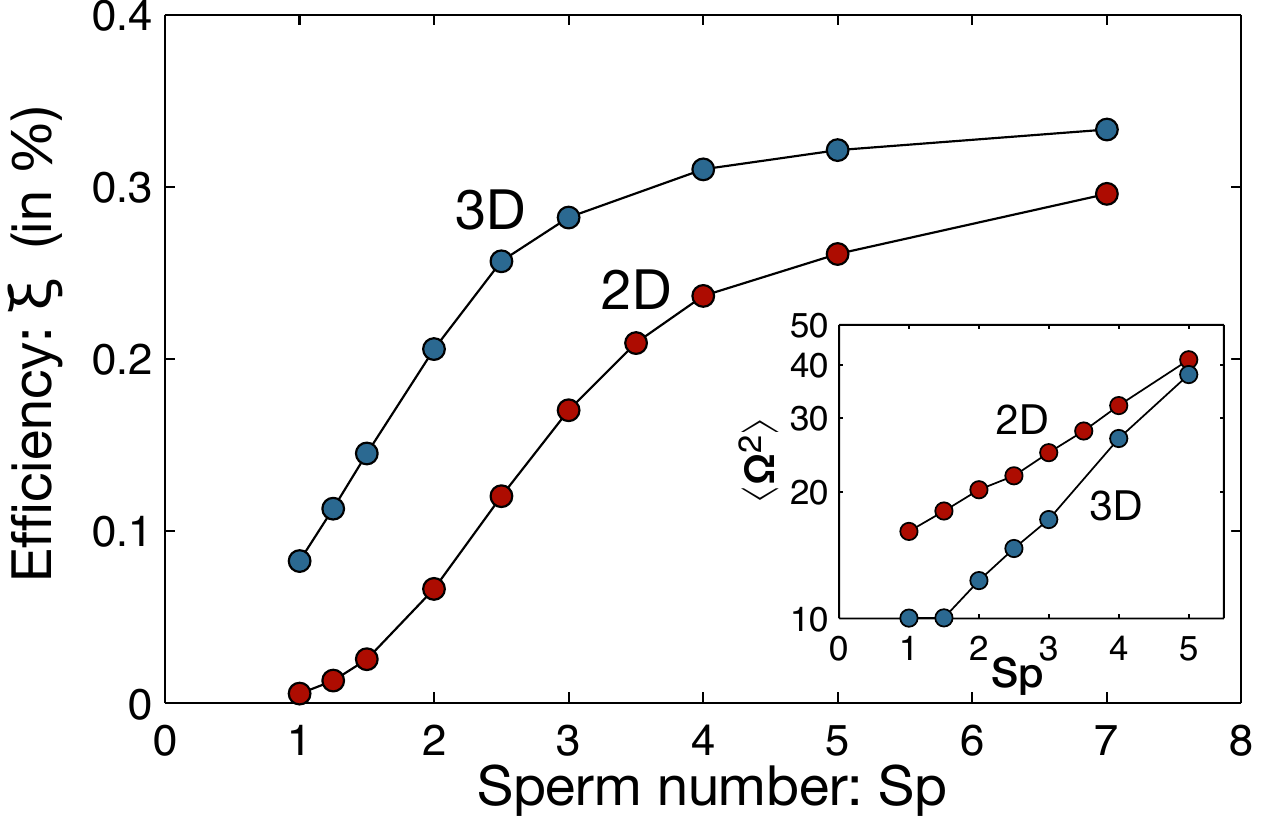} 
\caption{(color online) Optimal pumping efficiency, $\xi$, as a function of $\Sp$, for two (2D) and three-dimensional (3D) cilium kinematics.
The inset show the variation of the mean square curvature for the optimal kinematics (semilog scale). }
\label{fig:efficiency}
\end{figure}

The influence of the value of $\Sp$ on the optimal pumping efficiency is illustrated in  Fig.~\ref{fig:efficiency}, both for two and three-dimensional motions.  As expected, deformation in three dimensions is more efficient for all $\Sp$ since the number of degrees of freedom is larger, although the two cases converge to similar efficiencies for large $\Sp$.  In both  cases, the efficiency is a monotonically increasing function of the Sperm number: increasing $\Sp$ is equivalent to reducing the bending rigidity and thus allows larger curvatures for a lower energetic cost \footnote{Interestingly, although an increase in  bending rigidity generally increases the energetic cost, it is not always strictly true. In some cases, bending of the cilium  allows (elastic)  energy storage   to be restored later in the motion. For instance, the kinematics shown in Fig.~\ref{fig:3D}b (the $\Sp=2$ optimal) have maximum efficiency when $\Sp=3.2$ (although only  0.05\% more efficient than when $\Sp=\infty$).
}. 
In fact, the mean square curvature of the motion appears to be almost an exponential function of $\Sp$ in the range studied (inset of Fig.~\ref{fig:efficiency}). Unless an artificial dissipative term is introduced, one thus expects  the problem to become mathematically ill-posed in the limit of large $\Sp$, which is equivalent to considering only the energy dissipated in the fluid \cite{Osterman2011}.

\begin{figure}[b] 
\includegraphics[width=0.4\textwidth]{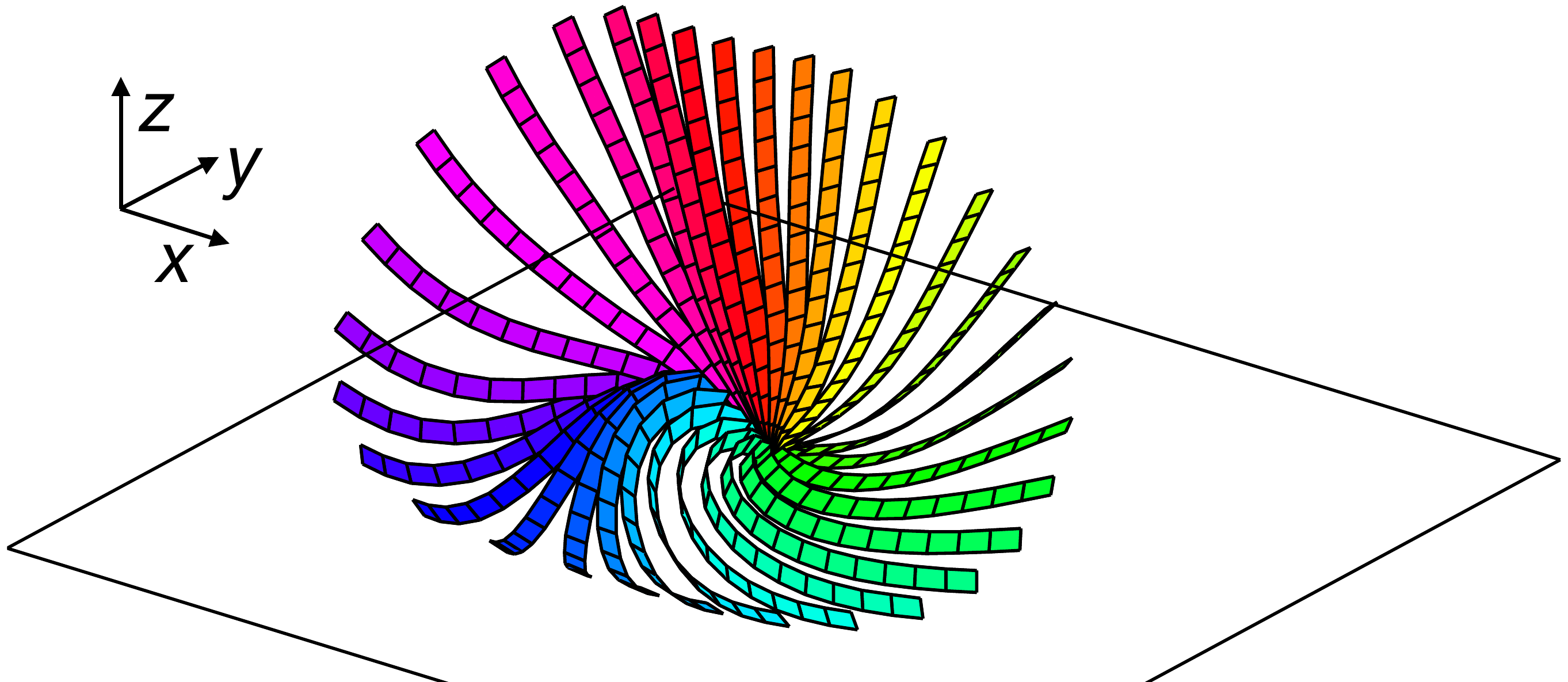} 
\caption{(color online) Illusion of twist for Sp=2. The three-dimensional kinematics is plotted as if the filament was a ribbon to emphasize the orientation of $\bd_1$.}
\label{fig:twist}
\end{figure}

As discussed above, we assumed here that the active internal moments cannot  produce torsion. But, even when torsion is not present, the filament kinematics can give the illusion of twist since its extremity rotates as shown in Fig.~\ref{fig:twist}. This is a classical result of differential geometry  \cite{Powers2010,Audoly2010} and it could explain why some studies reported the presence of twist on the cilia of \emph{Paramecium} \cite{Omoto1980}.

In summary, we have proposed in this Letter that, in order to derive the appropriate efficiency of cilia-driven fluid transport, the detailed internal structure of cilia has to be considered, and  energetic costs have to be calculated as the sum of the positive works done by the internal moments. 
Using this approach, we have developed a numerical model that allows to compute the kinematics of a wall-bound elastic cilium  transporting the surrounding fluid at minimum energetic cost. 
The optimal motions of the cilium  have been found to strongly depend on its  bending rigidity through a single dimensionless parameter, $\Sp$. These optimal kinematics were found to display the experimentally-observed two-stroke cycle, both in two and three dimensions.  Although  we have focused our study on the case of a single filament, cilia in biology are generally  densely packed on surfaces, and as such are strongly influenced by hydrodynamical interactions with their neighbors. These interactions are an intriguing  avenue for future work: they could affect flow transport and be responsible for the different cilium velocities observed during effective and recovery strokes \cite{Osterman2011}. 

\begin{acknowledgments}
We  acknowledge supports from the European Union (fellowship PIOF-GA-2009-252542 to C.E.) and the NSF (grant CBET-0746285 to E.L.).
\end{acknowledgments}

\bibliography{biblio}
\end{document}